\documentstyle[12pt,preprint,eqsecnum,aps]{revtex}
\begin{document}
\preprint{TRI-PP-94-69}
\title{ \bf
\centerline{An Effective Chiral Meson Lagrangian at $O(p^6)$}
\centerline{from the NJL Model} }
\author{
A.A.Bel'kov${}^1$,
A.V.Lanyov${}^1$,
A.Schaale${}^{2}$
\thanks{ supported by Deutscher Akademischer Austauschdienst, DAAD.},
S.Scherer${}^{2}$
\thanks{Address after Sept.\ 1, 1994:
Institut f\"ur Kernphysik, Johannes Gutenberg--Universit\"at,
J.\ J.\ Becher--Weg 45, D--55099 Mainz, Germany.} }
\address{
${}^1$  Particle Physics Laboratory, Joint Institute for Nuclear
        Research,
\\      141980 Dubna, Moscow Region, Russia
\\ ${}^2$ TRIUMF, 4004 Wesbrook Mall, Vancouver, B.\ C.,
\\ Canada V6T 2A3 }
\date{\today}
\maketitle
\begin{abstract}
In this work we present a strong chiral meson Lagrangian up to and
including $O(p^6)$ in the momentum expansion. It is derived from the
Nambu--Jona-Lasinio (NJL) model using the heat-kernel method.
Identities related to the properties of covariant derivatives of the
chiral matrix $U$ as well as field transformations have been used to
obtain a minimal set of linearly independent terms.
\end{abstract}
\pacs{PACS}

\section{\bf The Bosonization of the NJL Model}

The effective four-quark interaction of the NJL model \cite{njl} is a
low-energy approximation of QCD, the standard model of the
strong interactions of quarks and gluons.
The bosonization of the NJL model generates an effective chiral meson
Lagrangian which results from the quark determinant (see
\cite{volkov-NJL}-\cite{bijnens} and references therein).
Applying the heat-kernel techniques \cite{heat-other}-\cite{heat-our}
for the analytical calculation of the quark determinant one can derive
a momentum expansion of the effective meson Lagrangian.
In particular, the terms of $O(p^2)$ lead to the kinetic and mass
parts of the Lagrangian, and the terms of $O(p^4)$ can be brought into
the general form which was introduced by Gasser and Leutwyler
\cite{gasser}. A phenomenological analysis of the chiral coefficients
$L_i$ shows a good agreement with the predictions of the NJL model.
It is reasonable to expect the same of the next order in the
momentum expansion, where precise experimental data are not yet
available.

In previous works \cite{heat-our,our-boso} we have presented the
heat-kernel expansion of the quark determinant up to the order $h_6$ of
the heat coefficients, containing the complete information about
the $O(p^6)$ terms of the effective meson Lagrangian in the NJL model.
As far as precision experiments are becoming more sensitive, it will be
possible to observe effects which are related to the higher order of the
momentum expansion (see for example \cite{kaloshin} and \cite{barr}).
Here we derive the $O(p^6)$-Lagrangian from the NJL model.
A systematic study of the most general chiral
Lagrangian at $O(p^6)$ can be found in \cite{fearing-scherer}.

The starting point of our consideration is the effective four-quark
interaction of the NJL model with the Lagrangian
\begin{equation}
{\cal L}_{NJL} =  \overline{q}(i\widehat{\partial }-m_{0})q
                +{\cal L}_{int}\,,
\label{njl-lagr}
\end{equation}
where
$$
{\cal L}_{int} = 2G_{1}\bigg\{
   \bigg(\overline{q}\frac{\lambda^{a}}{2}q \bigg)^{2}
  +\bigg(\overline{q}i\gamma^{5}\frac{\lambda^{a}}{2}q \bigg)^{2}
                       \bigg\}
                -2G_{2}\bigg\{
   \bigg(\overline{q}\gamma^\mu \frac{\lambda^{a}}{2}q\bigg)^{2}
  +\bigg(\overline{q}\gamma^\mu \gamma^{5}\frac{\lambda^{a}}{2}q\bigg)^{2}
                       \bigg\}\,.
$$
Here $G_{1}$ and $G_{2}$ are empirical coupling constants,
$m_0$ is the current quark mass matrix, and $\lambda^a$ are the
generators of the $SU(n)$ flavor group, normalized according to
$\,\mbox{tr} (\lambda^{a} \lambda^{b}) = 2\delta^{ab}$.
Using a standard quark bosonization approach based on path
integral techniques one can derive an effective meson
action from the NJL Lagrangian (\ref{njl-lagr}).

First, one has to introduce colorless collective fields for the scalar
$(S)$, pseudoscalar $(P)$, vector $(V)$ and axial-vector $(A)$ mesons
associated with the quark bilinears,
$$
   S^{a}=-4G_{1}\overline{q}\frac{\lambda^{a}}{2}q\,,\enspace
   P^{a}=-4G_{1}\overline{q}i\gamma^{5}\frac{\lambda^{a}}{2}q\,, \enspace
   V_\mu^{a}=i4G_{2}\overline{q}\gamma_\mu \frac{\lambda^{a}}{2}q\,,\enspace
   A_\mu^{a}=i4G_{2}\overline{q}\gamma_\mu \gamma^{5}\frac{\lambda^{a}}{2}q\,,
$$
where $a$ is a flavor index.
After substituting these expressions into ${\cal L}_{NJL}$ the
interaction part of the Lagrangian becomes bilinear in the quark fields:
$$
{\cal L} =  \overline{q} i{\bf \widehat{D}} q
$$
with ${\bf \widehat{D}}$ being the Dirac operator in the presence of
the collective meson fields,
$$
i{\bf \widehat{D}}
        =  [i(\widehat{\partial} +\widehat{A}_R)
         - (\Phi    +m_0)] P_R
         + [i(\widehat{\partial} +\widehat{A}_L)
         - (\Phi^\dagger+m_0)] P_L.
$$
    Here $\Phi = S + iP$, $\widehat{V} = V_{\mu} \gamma^{\mu}$,
$\widehat{A} = A_{\mu} \gamma^{\mu}$;
$P_{R/L}={1 \over 2}(1 \pm \gamma_5)$ are chiral projectors;
$\widehat{A}^{R/L} = \widehat{V} \pm \widehat{A}$ are right and left
combinations of fields, and
$$
    S=S^a\frac{\lambda^a}{2} , \quad
    P=P^a\frac{\lambda^a}{2} , \quad
    V_{\mu}=-iV_{\mu}^a \frac{\lambda^a}{2} ,\quad
    A_{\mu}=-iA_{\mu}^a \frac{\lambda^a}{2}
$$
are the matrix-valued collective fields.

After integration over the quark fields the generating functional,
corresponding to the effective action of the NJL model for
collective meson fields, can be presented in the following form:
$$
{\cal Z} = \int {\cal D}\Phi\,{\cal D}\Phi^\dagger\,{\cal D}V\,{\cal D}A
           \,\,\mbox{exp}\big[ i{\cal S}(\Phi,\Phi^\dagger,V,A)\big]\,,
$$
where
\begin{equation}
{\cal S}(\Phi,\Phi^\dagger,V,A) =
          \int d^4 x
          \Big[-{1 \over{4G_1}} \mbox{tr} (\Phi^\dagger \Phi)
           - {1 \over{4G_2}} \mbox{tr} (V_\mu^2 + A_\mu^2)\Big]
   -i\,\,\mbox{Tr}^{\prime}\, [ \log( i \widehat{\bf D} ) ]
\label{action}
\end{equation}
is the effective action for scalar, pseudoscalar, vector and
axial-vector mesons.
The first term in Eq. (\ref{action}), quadratic in the meson fields,
arises from the linearization of the four-quark interaction.
The second one is the quark determinant describing the
interaction of mesons.
The quark determinant can be evaluated using the heat-kernel technique
with proper-time regularization
\cite{ebert-reinhardt,heat-other,heat-our}.
Then, the real part of $\log \big(\mbox{det}\,i{\bf \widehat{D}}\big)$
contributes to the even intrinsic parity part of the effective Lagrangian
while the imaginary part gives the odd intrinsic parity effective
Lagrangian which at $O(p^4)$ is related to the anomalous action of
Wess and Zumino \cite{wz,bijnens2}.

The logarithm of the modulus of the quark determinant is defined
in ``proper-time'' regularization as
\begin{eqnarray}
\log |\det i{\bf \widehat{D}}| \,  =
 -\, {1\over 2} \,\mbox{Tr}^{\prime}\, \log({\bf \widehat{D}}^\dagger
     {\bf \widehat{D}})
=
 -\,{1\over 2} \int^{\infty }_{1/\Lambda^2} d \tau \,\frac{1}{\tau}
  \,\mbox{Tr}^{\prime}\, \mbox{exp} \big(-{\bf\widehat{D}}^\dagger
                        {\bf\widehat{D}}\tau\big)
\label{proper-time}
\end{eqnarray}
with $\Lambda$ being the intrinsic regularization parameter.
The ''trace'' $\,\mbox{Tr}^{\prime}\,$ is to be understood as a
space-time integration and a ``normal'' trace with respect to
Dirac, color and flavor matrices,
$\mbox{Tr}^{\prime}=\int d^4 x \,\mbox{Tr}$, and
$\mbox{Tr}=\mbox{tr}\,_{\gamma} \cdot \,\mbox{tr}\,_C \cdot \,\mbox{tr}_F$.
The main idea of the heat-kernel method is to expand
$<x\mid \exp (-{\bf \widehat{D}}^\dagger {\bf \widehat{D}\tau})\mid y>$
around its ''free'' part
$$
<x\mid \exp (-(\hbox{\hbox{\vrule\vbox{\hrule\phantom{o}\hrule}\vrule}}
+\mu^{2})\tau)\mid y> =
{1\over (4\pi \tau)^{2}} e^{-\mu^{2}\tau+(x-y)^{2}/(4\tau)}
$$
in powers of the proper time $\tau$ with the so-called Seeley-deWitt
coefficients $h_{k}(x,y)$
$$
<x\mid \exp (-{\bf \widehat{D}}^\dagger {\bf \widehat{D}\tau})\mid y>
= {1\over (4\pi \tau)^{2}} e^{-\mu^{2}\tau+(x-y)^{2}/(4\tau)}
  \sum^{}_{k} h_{k}(x,y)\cdot \tau^{k}.
$$
The new mass scale $\mu$ arises as a nonvanishing
vacuum expectation value of the scalar field $S$, and corresponds to
the constituent quark mass.

After integration over $\tau$ in (\ref{proper-time}) one gets
\begin{equation}
\log |\det i{\bf \widehat{D}}| \,  =
 -\, {1\over 2} \frac{\mu^4}{(4 \pi)^2} \sum \limits_{k}
                \frac{\Gamma (k-2, \mu^2 / \Lambda^2)}{\mu^{2k}}
                \,\mbox{Tr}\, h_k,
\label{logarithm}
\end{equation}
where
$\Gamma (n,x)=\int^{\infty }_{x} d t \, e^{-t}t^{n-1}$
is the incomplete gamma function.
Using the definition of the gamma function $\Gamma (\alpha,x)$,
one can separate the divergent and finite parts of the quark
determinant
$$
{1\over 2}\log (\det {\bf \widehat{D}}^\dagger {\bf \widehat{D}})
   = B_{\hbox{pol}}+ B_{\hbox{log}}+ B_{\hbox{fin}},
$$
where
$$
B_{\mbox{pol}} = {1\over 2} {e^{-x} \over (4\pi )^{2}}
                 \left[- {\mu^{4}\over 2x^{2}} \,\mbox{Tr}\, h_{0}
                +{1\over x}({\mu^{4}\over 2} \,\mbox{Tr}\, h_{0}
                -\mu^{2}\,\mbox{Tr}\, h_{1} )\right]
$$
has a pole at $x=\mu^2 / \Lambda^2=0$,
$$
B_{\hbox{log}} = -{1\over 2} {1\over (4\pi )^{2}}\Gamma (0,x)
         \left[ {1\over 2}\mu^{4}\,\mbox{Tr}\, h_{0}-\mu^{2}\,\mbox{Tr}\, h_{1}
                 +\,\mbox{Tr}\, h_{2} \right]
$$
is logarithmically divergent, and the finite part has the form
$$
B_{\hbox{fin}} = - {1\over 2} {1\over (4\pi )^{2}}
                   \sum_{k=3}^{\infty} \mu^{4-2k}\Gamma (k-2,x)\,\mbox{Tr}\,
                   h_{k}.
$$

The very lengthy calculations of the Seeley-deWitt coefficients
$h_k$ can be only performed by computer support.
The calculation of the heat-coefficients is a recursive
process which can conveniently be done by Computer Algebra Systems
such as FORM and REDUCE.
In ref.\cite{heat-our} we have calculated the coefficients
up to the order $k=6$.
After voluminous computations one gets the expressions
for heat-coefficients $h_1, \ldots h_6$ up to terms of $O(p^6)$
(terms contributing only at higher orders are dropped)
\begin{eqnarray*}
h_{0} &=& 1 ,\\
h_{1} &=& -a ,\\
\,\mbox{Tr}\, h_{2} &=& \,\mbox{Tr}\, \bigg\{
   {1\over 12}(\Gamma_{\mu\nu})^{2} + {1\over 2}a^{2}\bigg\},
\\
\,\mbox{Tr}\, h_{3} &=& - {1\over 12} \,\mbox{Tr}\, \bigg\{
  {2a^{3}
 - S_\mu S^\mu
 + a(\Gamma_{\mu\nu})^{2}
 - {2\over 45}(\Gamma_{\alpha\beta\gamma })^{2}
 - {1\over 9}(\Gamma^\alpha {}_{\alpha\beta})^{2}}
 - {2\over 45}\Gamma_{\mu\nu}\Gamma^{\nu\alpha}\Gamma_{\alpha}{}^{\mu}
  \bigg\},
\\
\,\mbox{Tr}\, h_{4} &=& \,\mbox{Tr}\, \bigg\{
   {1\over 24}a^{4}
 + {1\over 12} \left( a^{2} S^\mu {}_\mu + a S_\mu S^\mu \right)
 + {1\over 720} \left( 7(S^\mu {}_\mu )^{2} -(S_{\mu\nu})^{2} \right)
\\ &&
 + {1\over 30}a^{2}(\Gamma_{\mu\nu})^{2}
 + {1\over 120}(a\Gamma_{\mu\nu})^{2}
 + {1\over 144}a \big[\Gamma^\mu {}_{\mu\nu},S^\nu \big]
 + {1\over 40}a \bigg(\Gamma_{\mu\nu}S^{\mu\nu}
 +                    {11\over 9}S_{\mu\nu}\Gamma^{\mu\nu} \bigg)
 \,\bigg\},
\\
\,\mbox{Tr}\, h_{5} &=&
 - \,\mbox{Tr}\, \bigg\{ {1\over 120} a^2 ( a^3 - 3S_\mu S^\mu)
 - {1\over 60}(a S_\mu)^2\bigg\},
\\
\,\mbox{Tr}\, h_{6} &=& {1\over 720} \,\mbox{Tr}\,  a^6 \, .
\end{eqnarray*}
    Here
$$
 \Gamma_{\mu\nu} = [d_\mu ,d_\nu]\,,\;\;
 \Gamma_{\lambda\mu\nu} = [d_{\lambda},\Gamma_{\mu\nu}]\,,\;\;
  S_\mu  = [d_\mu, a]\,,\;\;  S_{\mu\nu} = [d_\mu, S_\nu ]
$$
are commutators of the operators $d_\mu$ and $a$ which are defined by
the relations
$$
d_\mu = \partial_\mu + \Gamma_\mu, \quad
\Gamma_\mu = V_\mu + A_\mu \gamma^{5}, \quad
a(x) = i\widehat{\nabla} H + H^\dagger H
      + \frac{1}{4}[\gamma^\mu , \gamma^\nu ]\Gamma_{\mu\nu} - \mu^2.
$$
Moreover,
$$
  H=P_{R}(\Phi+m_0) + P_{L}(\Phi^\dagger+m_0) = S+m_0+i\gamma_{5}P\,,
$$
and
$$
  \Gamma_{\mu\nu}=[d_\mu ,d_\nu ] =
  \partial_\mu \Gamma_\nu -\partial_\nu \Gamma_\mu
  +[\Gamma_\mu ,\Gamma_\nu ]=F^{V}_{\mu\nu}+\gamma^{5}F^{A}_{\mu\nu}\,,
$$
where $F^{V/A}_{\mu\nu}$ are the field strength tensors,
$$F^{V}_{\mu\nu}=
\partial_\mu V_\nu -\partial_\nu V_\mu
                   +[V_\mu ,V_\nu ]+[A_\mu ,A_\nu ]\,,\;\;
F^{A}_{\mu\nu}=
\partial_\mu A_\nu -\partial_\nu A_\mu
                   +[V_\mu ,A_\nu ]+[A_\mu ,V_\nu ]\,,
$$
and
$$
\nabla_\mu H=\partial_\mu H+[V_\mu ,H]-\gamma^{5}\{A_\mu ,H\}\,.
$$

\section{\bf The Chiral Lagrangian}

We will consider here a nonlinear parameterization of chiral symmetry
corresponding to the representation $ \Phi = \Omega \,\Sigma
\,\Omega$.
The matrix of scalar fields $\Sigma(x)$ belongs to the
diagonal flavor group, while the matrix $\Omega(x)$ represents the
pseudoscalar degrees of freedom $\varphi$ living in the coset space
$SU(n)_L \! \times \! SU(n)_R/SU_V(n)$. It can be parameterized by
the $SU(n)$ matrix
\begin{eqnarray*}
  \Omega (x) = \exp \left(\frac{i}{\sqrt{2}F_0} \varphi (x) \right)\,,
\quad
  \varphi (x) = \varphi^a(x)\frac{\lambda^a}{2}\,,
\end{eqnarray*}
with $F_0$ being the bare $\pi$ decay constant.
Under chiral rotations
\begin{eqnarray*}
q \rightarrow \widetilde{q} = \left( P_L \xi_L + P_R \xi_R \right)q
\end{eqnarray*}
the fields $\Phi$ and $A_\mu^{R/L}$ are transforming as
\begin{eqnarray*}
     \Phi  \rightarrow \widetilde{\Phi} = \xi_L \Phi \xi^\dagger_R\,,
\end{eqnarray*}
and
\begin{eqnarray}
A_\mu^R \rightarrow \widetilde{A}_\mu^R =
                        \xi_R ( \partial_\mu + A_\mu^R ) \xi^\dagger_R,
\quad
A_\mu^L \rightarrow \widetilde{A}_\mu^L =
                        \xi_L ( \partial_\mu + A_\mu^L ) \xi^\dagger_L.
\nonumber
\end{eqnarray}

The effective meson Lagrangian in terms of the collective
fields is obtained from the quark determinant by calculating
the trace over Dirac indices in $\,\mbox{Tr}\, h_i(x)$.
The ``divergent'' part of the effective meson Lagrangian is
defined by the coefficients $h_0, h_1$ and $h_2$ of the expansion
(\ref{logarithm})
\begin{eqnarray}
{\cal L}_{div} &=&
    \frac{N_c}{16 \pi^2} \mbox{tr} \bigg\{
    \Gamma \bigg( 0,\frac{\mu^2}{\Lambda^2} \bigg) \bigg[
    D^{\mu}(\Phi+m_0)\,\overline{D}_{\mu}(\Phi+m_0)^\dagger - {\chi}^2
    +\frac{1}{6}\,\bigg((F^L_{\mu \nu})^2 + (F^R_{\mu \nu})^2
                  \bigg) \bigg]
\nonumber
\\
&+& 2 \bigg[ \Lambda^2 e^{-\mu^2 / \Lambda^2}
    - \mu^2 \Gamma \bigg( 0,\frac{\mu^2}{\Lambda^2} \bigg) \bigg]
      {\chi} \bigg\},
\label{ap1}
\end{eqnarray}
    where ${\chi} = (\Phi+m_0)(\Phi+m_0)^\dagger - \mu^2$ and
$F^{R/L}_{\mu\nu} = F^V_{\mu\nu} \pm F^A_{\mu\nu}$.
The covariant derivatives $D_{\mu}$ and $\overline{D}_{\mu}$
are defined as
$$
D_{\mu}* = \partial_{\mu}* + (A^L_{\mu}* - *A^R_{\mu})\,,
\quad
\overline{D}_{\mu}* = \partial_{\mu}* + (A^R_{\mu}*-*A^L_{\mu})\,,
$$
where it is understood that $D_\mu$ and $\overline{D}_\mu$ act upon
expressions transforming as $\xi_L \cdots \xi_R^\dagger$ and
$\xi_R \cdots \xi_L^\dagger$, respectively.
Assuming $\Sigma\approx\mu$ and therefore
$\Phi=\mu \Omega^2 \equiv \mu U$,
the Lagrangian of $O(p^2)$ can be written in the form
$$
{\cal L}^{(2)}_{eff}=-\frac{F^2_0}{4} \,\mbox{tr}\,
                      \big( L_{\mu} L^{\mu} \big)
                     +\frac{F_0^2}{4} \,\mbox{tr}\,
                      \big( \chi U^\dagger + U\chi^\dagger \big)\,,
$$
where $L_{\mu}=D_{\mu}U\,U^\dagger$.
The bare constant $F_0$ and the matrix
$\chi  = \mbox{diag}\,(\chi^2_u,\chi^2_d,\chi^2_s)$ are given by
$F^2_0 = y N_c \mu^2 / (4 \pi^2)$ and
$ \chi^2_i=m^i_0\mu /(G_1 F^2_0)
= - 2m^i_0\!\!<\!\!\overline{q}q\!\!>\!\! F^{-2}_0$,
where $y=\Gamma\big(0,\mu^2/\Lambda^2\big)$ and
$<\overline{q} q >$ is the quark condensate.

The terms of $O(p^4)$ of the effective Lagrangian
result from the logarithmically divergent part of the quark determinant and
from the coefficients $h_3$ and $h_4$ contributing to the finite part.
Using properties of the derivatives (see appendix,
Eqs.(\ref{appendix1},\ref{appendix2}))
the finite contribution of $O(p^4)$ can be written as
\begin{eqnarray}
{\cal L}^{(p^4)}_{fin} &=&
    \frac{N_c}{32 \pi^2 \mu^4} \mbox{tr} \bigg\{
    \frac{1}{3}\,\Big[ \mu^2 D^2 (\Phi+m_0)\,\overline{D}^2(\Phi+m_0)^\dagger
                -\big(D^{\mu}(\Phi+m_0)\,\overline{D}_{\mu}(\Phi+m_0)^\dagger
       \big)^2\Big]
\nonumber \\
&+& \frac{1}{6} \big(D_{\mu}(\Phi+m_0)\,\overline{D}_{\nu}(\Phi+m_0)^\dagger
                  \big)^2
\nonumber \\
&-& \mu^2 \big( {\chi} D_{\mu}(\Phi+m_0)\,\overline{D}^{\mu}(\Phi+m_0)^\dagger
    +{\chi}^\dagger\,\overline{D}_{\mu} (\Phi+m_0)^\dagger D_{\mu}(\Phi+m_0)
          \big)
\nonumber \\
&+& \frac{2}{3}\mu^2\,\Big(
    D^{\mu}(\Phi+m_0)\,\overline{D}^{\nu}(\Phi+m_0)^\dagger \,F^L_{\mu \nu}
    +\overline{D}^{\mu}(\Phi+m_0)^\dagger \,D^{\nu}(\Phi+m_0)\,F^R_{\mu \nu}
                      \Big)
\nonumber \\
&+& \frac{1}{3}\mu^2 F^R_{\mu \nu}(\Phi+m_0)^\dagger F^{L\,\mu \nu} (\Phi+m_0)
  - \frac{1}{6}\mu^4\,\Big[(F^L_{\mu \nu})^2 + (F^R_{\mu\nu})^2\Big]
    \bigg\}\,,
\label{ap2}
\end{eqnarray}
   where ${\chi}^\dagger=(\Phi+m_0)^\dagger (\Phi+m_0)-\mu^2$.
   We will assume the approximation
$\Gamma ( k,\mu^2 / \Lambda^2 ) \approx \Gamma(k)$, valid for
$k \geq 1$, and $\mu^2 / \Lambda^2 \ll 1$.

The effective meson Lagrangian of $O(p^4)$, Eq.(\ref{ap2}), can
be brought into the standard form introduced by Gasser and Leutwyler
in ref.\cite{gasser} (see appendix, Eq.(\ref{l42}))
After using the field transformations \cite{fields}, which are at $O(p^4)$
equivalent to the application of the classical equation of motion (EOM)
(see appendix \ref{eom2}), the NJL model gives the following
predictions for the chiral coefficients
\begin{eqnarray}
&&
L_1=\frac{N_c}{16 \pi^2}\frac{1}{24}\,,\quad
L_2= \frac{N_c}{16 \pi^2}\frac{1}{12}\,,\quad
L_3= -\,\frac{N_c}{16 \pi^2}\frac{1}{6}\,,\quad
\nonumber \\
&&
L_4=0 \,,\quad L_5= \frac{N_c}{16 \pi^2}x(y-1)\,,\quad L_6 = 0\,,
\nonumber \\&&
L_7= -\,\frac{N_c}{16\pi^2}\frac{1}{6}\bigg(xy-\frac{1}{12}\bigg)\,,\quad
L_8= \frac{N_c}{16 \pi^2}\bigg[
       \bigg(\frac{1}{2}x-x^2\bigg)y-\frac{1}{24}\bigg]\,,\quad
\nonumber \\
&&
L_9=  \frac{N_c}{16 \pi^2}\frac{1}{3}\,,\quad
L_{10}=  -\, \frac{N_c}{16 \pi^2}\frac{1}{6}\,,
\nonumber \\ &&
H_1=-\,\frac{N_c}{16 \pi^2}\frac{1}{6} \bigg( y-\frac{1}{2} \bigg)\,,\quad
H_2= -\,\frac{N_c}{16 \pi^2}\bigg[ (x+2x^2)y-\frac{1}{12} \bigg] \,,
\label{Lcoeff}
\end{eqnarray}
where $x = -\mu F_0^2/(2 <\! \! \overline{q} q \! \! >)$ and
$y = 4\pi^2F_0^2/(N_c \mu^2)$.
\vspace{0.3cm}

Analogous as for the $O(p^4)$ Lagrangian we present the
$p^6$ Lagrangians in a ''minimal'' form, avoiding redundant terms.
The identities and relations which we
have used in order to keep the number of terms as small as possible
can be found in the appendix.
It is important to realize that the field transformations used to
bring the Lagrangian of $O(p^4)$ into the form of Gasser and Leutwyler
also result in contributions at $O(p^6)$ and higher \cite{fields}.
Furthermore, we eliminated (see app.) terms at $O(p^6)$ using field
transformations.

The final effective $p^6$-Lagrangian has the form (see appendix for details)

\begin{eqnarray}
%
{\cal L}_{eff}^{(6)} &=& \frac{N_c}{32\pi^2\mu^2}
          \mbox{tr} \bigg\{
         -\frac{1}{10} \big( L_\mu L_\nu L^\nu \big)^2
 \\ &&
         +\frac{5}{18} \big( L_\mu L^\mu \big)^3
 \\ &&
         -\frac{1}{45} L_\alpha L^\alpha \big(L_\mu L_\nu\big)^2
 \\ &&
         +\frac{1}{30} \big( L_\mu L_\nu L_\alpha \big)^2
 \\ &&
         -\frac{1}{10} \big( L_\mu L_\nu L^\mu \big)^2
 \\ &&
         +\frac{1}{30}\Big(
          D_{\mu}U\,\overline{D}_{\nu}U^\dagger
          D_{\alpha}D^{\nu}U\,\overline{D}^{\alpha}\overline{D}^{\mu}U^\dagger
         +\overline{D}_{\mu}U^\dagger D_{\nu}U
          \overline{D}_{\alpha}\overline{D}^{\nu}U^\dagger\,D^{\alpha}D^{\mu}U
               \Big)
 \\ &&
         -\frac{1}{30}\Big(
          D_{\mu}U\,\overline{D}_{\nu}U^\dagger
          D_{\alpha}D^{\mu}U\,\overline{D}^{\alpha}\overline{D}^{\nu}U^\dagger
         +\overline{D}_{\mu}U^\dagger D_{\nu}U
          \overline{D}_{\alpha}\overline{D}^{\mu}U^\dagger\,D^{\alpha}D^{\nu}U
               \Big)
 \\ &&
         +\Big[4c + \frac{1}{2} \Big( \frac{73}{90}-x \Big)\Big]
          D_{\mu}U\,\overline{D}_{\nu}U^\dagger
          D^{\nu}U\,\overline{D}^{\mu}U^\dagger (\chi U^\dagger+U\chi^\dagger)
 \\ &&
         +\Big[2c - \frac{47}{180}\Big]
          \big(D_{\mu}U \overline{D}_{\nu}U^\dagger\big)^2
          (\chi U^\dagger+U\chi^\dagger)
 \\ &&
         +\Big[-2c + \frac{1}{6}\Big(x+\frac{1}{15}\Big)\Big]
          \big(D_{\mu}U \overline{D}^{\mu}U^\dagger\big)^2
          (\chi U^\dagger+U\chi^\dagger)
 \\ &&
         +\Big[ c+\frac{1}{60}\Big] \Big(
            \,\chi (
      \overline{D}^\mu U^\dagger \, D_\mu D_\nu U \, \overline{D}^\nu U^\dagger
    + \overline{D}^\nu U^\dagger \, D_\mu D_\nu U \, \overline{D}^\mu U^\dagger
             )
\nonumber \\ &&
\quad\quad\quad\quad \;
           +\chi^\dagger(
              D^\mu U \, \overline{D}_\mu \overline{D}_\nu U^\dagger \, D^\nu U
            + D^\nu U \, \overline{D}_\mu \overline{D}_\nu U^\dagger \, D^\mu U
                     )
          \Big)
 \\ &&
         +\Big[ 3c - \frac{1}{3} \Big(x-\frac{1}{20}\Big) \Big]
           \Big(
            \chi (  \overline{D}_\mu\overline{D}_\nu U^\dagger D^\mu U
                \overline{D}^\nu U^\dagger
              + \overline{D}^\nu U^\dagger D^\mu U
                \overline{D}_\mu \overline{D}_\nu U^\dagger
             )
\nonumber \\ &&
\quad\quad\quad\quad\quad\quad\quad \;\;\;\;\,
           +\chi^\dagger(D_\mu D_\nu U \overline{D}^\mu U^\dagger D^\nu U
                     +D^\nu U \overline{D}^\mu U^\dagger D_\mu D_\nu U)
          \Big)
 \\ &&
         +\Big[ 2c(1-6x)

+\frac{1}{3}\Big(2x^2-\frac{1}{2}x+\frac{1}{20}\Big)+2\tilde{c}\Big]
          \chi^\dagger U \overline{D}_{\mu}U^\dagger \chi U^\dagger D^{\mu}U
 \\ &&
         +\Big[c(12x+1)-\frac{1}{3}\Big(4x^2
              +\frac{1}{4}x-\frac{1}{40}\Big)  -2\tilde{c}\Big]
         \Big( \chi^\dagger \chi  \overline{D}_{\mu}U^\dagger D^{\mu}U
              +\chi \chi^\dagger D_{\mu}U\,\overline{D}^{\mu}U^\dagger \Big)
 \\ &&
         +\Big[ c + \frac{x}{3} \Big(2x -\frac{1}{4}\Big) \Big]
          \Big( U\chi^\dagger U \chi^\dagger
                D_{\mu}U \overline{D}_{\mu}U^\dagger
               +U^\dagger \chi  U^\dagger \chi
                \overline{D}_{\mu}U^\dagger D^{\mu}U \Big)
 \\ &&
         +\Big[ c(1+6x) - \frac{x}{3} \Big( x+\frac{1}{4}\Big) -\tilde{c} \Big]
          \Big((\chi  \overline{D}_{\mu}U^\dagger)^2
         + (\chi^\dagger D_{\mu}U )^2 \Big)
 \\ &&
         + \Big[x \Big( - 6c + x -\frac{1}{12}\Big) +\tilde{c}\Big]
          \Big(D^{\prime}_\mu(\chi U^\dagger+U\chi^\dagger)
          (\chi \overline{D}^\mu U^\dagger +D^\mu U \chi^\dagger )
\nonumber \\ &&
\quad\quad\quad\quad \quad\quad \quad\quad \quad\quad \;\;\;\,
          +\overline{D}^{\prime}_\mu (\chi^\dagger U+U^\dagger \chi )
          (\chi^\dagger D^\mu U + \overline{D}^\mu U^\dagger \chi ) \Big)
 \\ &&
         +\frac{x}{12}\Big(
          D^{\prime 2}(\chi U^\dagger+U\chi^\dagger)
          D_{\mu}U\,\overline{D}^{\mu}U^\dagger
         +\overline{D}^{\prime2}(\chi^\dagger U+U^\dagger \chi ) \,
          \overline{D}_{\mu}U^\dagger D^{\mu}U
                      \Big)
 \\ &&
         \Big[-6cxy + \tilde{c} \Big]\Big(
          D_{\mu}(\chi-U\chi^\dagger U ) \, \overline{D}^{\mu}\chi^\dagger
          +\overline{D}_{\mu}(\chi^\dagger-U^\dagger \chi  U^\dagger) \,
D^{\mu}\chi
            \Big)
 \\ &&
         -cD_{\mu}(U\chi^\dagger U-\chi )
            \overline{D}_{\mu}(\chi^\dagger-U^\dagger \chi  U^\dagger)
 \\ &&
         -\frac{9}{2}c^2y\Big( D_{\mu}^{\prime}(\chi
U^\dagger-U\chi^\dagger)\Big)^2
 \\ &&
         -\frac{9}{2}c^2y D_{\mu}^{\prime}(\chi U^\dagger-U\chi^\dagger)
                     [(\chi U^\dagger-U\chi^\dagger),D^{\mu}U\cdot U^\dagger ]
 \\ &&
        -\frac{x^2}{6}\Big(
          \big( D^{\prime}_{\mu}(\chi U^\dagger+U\chi^\dagger) \big)^2
         +\big(\overline{D}^{\prime}_{\mu} (\chi^\dagger U+U^\dagger \chi )
\big)^2
                     \Big)
 \\ &&
         + \Big[ \frac{9}{4}yc^2+ 6cx(1 - 2xy)
                  +\frac{1}{3}x\Big(2x^2 -3x+\frac{1}{8} \Big) \Big]
          \Big((\chi  U^\dagger)^3 + (\chi^\dagger U)^3 \Big)
 \\ &&
         + \Big[-\frac{9}{4}c^2y - 6cx(1-2xy) + 2x^3 +x^2 -\frac{1}{24}x \Big]
          ( U^\dagger \chi  \chi^\dagger \chi  + U \chi^\dagger \chi
\chi^\dagger )
 \\[0.9em] &&
        -\frac{1}{45} \Big(F^L_{\mu \nu}
          \big\{ D_{\alpha}U\,\overline{D}^{\alpha}U^\dagger ,
                D^{\mu}U\,\overline{D}^{\nu}U^\dagger \big\}
                            +F^R_{\mu \nu}
          \big\{ \overline{D}_{\alpha}U^\dagger D^{\alpha}U,
                 \overline{D}^{\mu}U^\dagger D^{\nu}U \big\} \Big)
 \\ &&
        +\frac{5}{36}
         \Big(\,F^L_{\mu \nu} \big( D^{\mu}U\,\overline{D}_{\alpha}U^\dagger
                                  D^{\nu}U\,\overline{D}^{\alpha}U^\dagger
                                 +D_{\alpha}U\,\overline{D}^{\mu}U^\dagger
                                  D^{\alpha}U\,\overline{D}^{\nu}U^\dagger
                            \big)
\nonumber \\ &&
\quad\quad   +F^R_{\mu \nu} \big( \overline{D}^{\mu}U^\dagger\,D_{\alpha}U
                                  \overline{D}^{\nu}U^\dagger\,D^{\alpha}U
                                 +\overline{D}_{\alpha}U^\dagger\,D^{\mu}U
                                  \overline{D}^{\alpha}U^\dagger\,D^{\nu}U
                            \big)
         \Big)
 \\ &&
        -\frac{3}{20}
         \big(F^L_{\mu \nu} D_{\alpha}U\,\overline{D}^{\mu}U^\dagger
                            D^{\nu}U\,\overline{D}^{\alpha}U^\dagger
             +F^R_{\mu \nu} \overline{D}_{\alpha}U^\dagger\,D^{\mu}U
                            \overline{D}^{\nu}U^\dagger\,D^{\alpha}U
         \big)
 \\ &&
        -\frac{7}{15}
         \big(F^L_{\mu \nu} D^{\mu}U\,\overline{D}_{\alpha}U^\dagger
                            D^{\alpha}U\,\overline{D}^{\nu}U^\dagger
             +F^R_{\mu \nu} \overline{D}^{\mu}U^\dagger\,D_{\alpha}U
                            \overline{D}^{\alpha}U^\dagger\,D^{\nu}U
         \big)
 \\ &&
        -\frac{2}{15} \Big(F^{L\,\alpha}{}_{\alpha\nu}
          U\overline{D}_{\mu}U^\dagger D^{\nu}U \overline{D}^{\mu}U^\dagger
                           +F^{R\,\alpha}{}_{\alpha\nu}
          U^\dagger D_{\mu}U \overline{D}^{\nu}U^\dagger D^{\mu}U
                       \Big)
 \\ &&
        +\frac{1}{5} \Big(F^{L\,\alpha}{}_{\alpha \nu}U
         \big(\overline{D}^{\nu}U^\dagger D_{\mu}U\,\overline{D}^{\mu}U^\dagger
         +\overline{D}_{\mu}U^\dagger D^{\mu}U \overline{D}^{\nu}U^\dagger\big)
\nonumber \\ &&
        \quad \;  +F^{R\,\alpha}{}_{\alpha \nu}U^\dagger
          \big( D^{\nu}U \overline{D}_{\mu}U^\dagger D^{\mu}U
         +D_{\mu}U \overline{D}^{\mu}U^\dagger D^{\nu}U \big) \Big)
 \\ &&
        -\frac{x}{6}\Big(
          (\chi U^\dagger+U\chi^\dagger) \big\{ F^L_{\mu \nu},
                      D^{\mu}U\,\overline{D}^{\nu}U^\dagger \big\}
         +(\chi^\dagger U+U^\dagger \chi ) \big\{ F^R_{\mu \nu},
                      \overline{D}^{\mu}U^\dagger D^{\nu}U\big\}
                     \Big)
 \\ && 
        +\Big[ c + \frac{1}{60} \Big] \Big(
          (\chi U^\dagger-U\chi^\dagger) \big[ F^L_{\mu \nu},
                      D^{\mu}U\,\overline{D}^{\nu}U^\dagger \big]
         +(\chi^\dagger U-U^\dagger \chi ) \big[ F^R_{\mu \nu},
                      \overline{D}^{\mu}U^\dagger D^{\nu}U\big]
                     \Big)
 \\ &&
         + \Big[ 2c -\frac{1}{2}x \Big]
          \Big(\chi (\overline{D}^\mu \overline{D}^\nu U^\dagger \,
F^L_{\mu\nu}
                 -F^R_{\mu\nu} \overline{D}^\mu \overline{D}^\nu U^\dagger)
\nonumber \\ &&
\quad\quad\quad\quad\;\;
               + \chi^\dagger (D^\mu D^\nu U F^R_{\mu\nu}
                            - F^L_{\mu\nu} D^\mu D^\nu U) \Big)
 \\ &&
         + \Big[ 2c - \frac{1}{6}x \Big]
          \Big( \chi ( U^\dagger F^L_{\mu\nu} \, D_\mu D_\nu U\,U^\dagger
                  -U^\dagger D_\mu D_\nu U \, F^R_{\mu\nu} U^\dagger)
\nonumber \\ &&
\quad\quad\quad\quad \;\;\;
               +\chi^\dagger( U F^R_{\mu\nu} \,
                           \overline{D}_\mu \overline{D}_\nu U^\dagger \, U
                          -U \overline{D}_\mu \overline{D}_\nu U^\dagger \,
                           F^L_{\mu\nu} U)
          \Big)
 \\ &&
         + \Big[ -2c +\frac{1}{2}x +\frac{1}{20} \Big]
          \Big( F^{L\,\alpha}{}_{\alpha\mu}
                (D^\mu U \, \chi^\dagger - \chi  \overline{D}^\mu U^\dagger )
\nonumber \\ &&
\quad\quad \quad\quad \quad \quad \quad\;\;\;\;\;
               +F^{R\, \alpha}{}_{\alpha\mu}
                (\overline{D}^\mu U^\dagger \, \chi  - \chi^\dagger D^\mu U)
\Big)
 \\ &&
         + \Big[ 2c - \frac{1}{6}x -\frac{1}{20} \Big]
          \Big(
          F^{L\,\alpha}{}_{\alpha\mu}
          ( D^{\mu}U\, U^\dagger \chi  U^\dagger
           -U \chi^\dagger U \overline{D}^\mu U^\dagger)
\nonumber \\ &&
\quad\quad \quad\quad \quad\quad \;\;\;\;
         +F^{R\,\alpha}{}_{\alpha\mu}
          ( \overline{D}^{\mu}U^\dagger \, U \chi^\dagger U
           -U^\dagger \chi  U^\dagger D^\mu U )
          \Big)
 \\ &&
        -\frac{2}{15} \Big(
          F^{L\,\mu\alpha}{}_{\alpha\nu}
          \big(D_{\mu}U\overline{D}^{\nu}U^\dagger
              -D^{\nu}U \overline{D}_{\mu}U^\dagger\big)
         +F^{R\,\mu\alpha}{}_{\alpha\nu}
          \big(\overline{D}_{\mu}U^\dagger D^{\nu}U
         -\overline{D}^{\nu}U^\dagger D^{\mu}U \big)
         \Big)
 \\  &&
        -\frac{1}{180}F^L_{\mu\nu}D_{\alpha}U\,
                        F^{R\,\mu\nu}\overline{D}^{\alpha}U^\dagger
 \\ &&
        -\frac{1}{5} F^L_{\alpha\mu}
D_{\nu}U\,F^{R\,\alpha\nu}\overline{D}^{\mu}U^\dagger
 \\ &&
        -\frac{1}{5} F^L_{\alpha\mu}
D^{\mu}U\,F^{R\,\alpha\nu}\overline{D}_{\nu}U^\dagger
 \\ &&
        -\frac{83}{360}\Big(
          \big(F^L_{\mu\nu}\big)^2
          D_{\alpha}U\,\overline{D}^{\alpha}U^\dagger
         +\big(F^R_{\mu\nu}\big)^2
          \overline{D}_{\alpha}U^\dagger\,D^{\alpha}U \Big)
 \\ &&
        +\frac{1}{15}\Big(
          D_{\alpha}U\,\overline{D}^{\alpha}U^\dagger U
          F^R_{\mu\nu}U^\dagger F^{L\,\mu\nu}
         +\overline{D}_{\alpha}U^\dagger\,D^{\alpha}U\,U^\dagger
          F^L_{\mu\nu}UF^{R\,\mu\nu}
                      \Big)
 \\ &&
        +\frac{1}{2}\Big(
          F^L_{\mu\alpha}F^{L\,\alpha\nu}
          D^{\mu}U\,\overline{D}_{\nu}U^\dagger
         +F^R_{\mu\alpha}F^{R\,\alpha\nu}
          \overline{D}^{\mu}U^\dagger D_{\nu}U
                      \Big)
 \\ &&
        +\frac{11}{30}\Big(
          F^L_{\mu\alpha}F^{L\,\alpha\nu}
          \overline{D}_{\nu}U^\dagger D^{\mu}U
         +F^R_{\mu\alpha} F^{R\,\alpha\nu}
          \overline{D}_{\nu}U^\dagger D^{\mu}U
                      \Big)
 \\ &&
       +\frac{1}{6}\Big(
         F^L_{\mu\nu}\big( D^{\mu}D_{\alpha}U
        +D_{\alpha}D^{\mu}U\big)F^{R\,\nu\alpha}U^\dagger
        +F^R_{\mu\nu}\big( \overline{D}^{\mu}\overline{D}_{\alpha}U^\dagger
        +\overline{D}_{\alpha}\overline{D}^{\mu}U^\dagger\big)F^{L\,\nu\alpha}U
                      \Big)
 \\ &&
        -\frac{1}{6}\Big(
          F^{L\,\alpha}{}_{\alpha\nu} \big(
             UF^{R\,\mu\nu}\overline{D}_{\mu}U^\dagger
            +D_{\mu}UF^{R\,\mu\nu}U^\dagger \big)
\nonumber \\ && 
          \quad \;
         +F^{R\,\alpha}{}_{\alpha\nu} \big(
             U^\dagger F^{L\,\mu\nu}D_{\mu}U
            +\overline{D}_{\mu}U^\dagger F^{L\,\mu\nu}U \big)
                    \Big)
 \\ &&
        -\frac{1}{15}\Big(
          F^{L\,\alpha}{}_{\alpha\nu}
          [D_{\mu}U\,U^\dagger, F^{L\,\mu\nu}]
         +F^{R\,\alpha}{}_{\alpha\nu}
          [\overline{D}_{\mu}U^\dagger, UF^{R\,\mu\nu}]
                     \Big)
 \\ &&
        +\frac{1}{15}
          F^{L\,\mu}{}_{\mu\alpha}U F^R_{\nu}{}^{\nu\alpha}U^\dagger
 \\ &&
    +\frac{1}{6}(x-5)\Big((\chi U^\dagger+U\chi^\dagger)(F^L_{\mu\nu})^2
    +(\chi^\dagger U + U^\dagger \chi )(F^R_{\mu\nu})^2 \Big)
 \\ &&
         + \Big[ c - \frac{1}{6}\Big(x-\frac{1}{5}\Big) \Big]
          \big(\chi  U^\dagger F^L_{\mu\nu} U F^{R\,\mu\nu} U^\dagger
         +     \chi^\dagger U F^R_{\mu\nu} U^\dagger F^{L\,\mu\nu} U\big)
 \\ &&
         - \Big[ c - \frac{1}{6}\Big(x-\frac{1}{5}\Big) \Big]
          \big(\chi  F^R_{\mu\nu} U^\dagger F^{L\,\mu\nu}
         +     \chi^\dagger F^L_{\mu\nu} U F^{R\,\mu\nu}\big)
 \\[0.9em] &&
        +\frac{41}{540}\Big(
          \big( F^L_{\mu\nu\alpha} \big)^2
         +\big( F^R_{\mu\nu\alpha} \big)^2 \Big)
 \\ &&
        -\frac{7}{135}\Big(
          \big( F^{L\,\mu}{}_{\mu\alpha}\big)^2
         +\big( F^{R\,\mu}{}_{\mu\alpha}\big)^2 \Big)
 \\ &&
        +\frac{1}{3}\Big(
          F^L_{\mu\nu}F^{L\,\mu\alpha} F^{L\,\nu}{}_{\alpha}
         +F^R_{\mu\nu}F^{R\,\mu\alpha} F^{R\,\nu}{}_{\alpha} \Big)
                         \bigg\}
\\[0.9em] &+&
 \frac{N_c}{32\pi^2\mu^2} \,\mbox{tr}\,
           \big(\chi U^\dagger - U\chi^\dagger \big) \bigg\{
\nonumber \\ &&
       \frac{1}{180} \,\mbox{tr}\, \Big(
         (  D_\nu U \overline{D}_\mu U^\dagger
          + D_\mu U \overline{D}_\nu U^\dagger )
         (  D_\mu D_\nu U \,U^\dagger
          - U \overline{D}_\mu \overline{D}_\nu U^\dagger)
                         \Big)
 \\ &&
     + \Big[ c \Big(\frac{1}{3}+2x-2xy\Big)
             -\frac{1}{18}\Big(x-\frac{1}{10}\Big)+\frac{1}{3}\tilde{c}\Big]
       \,\mbox{tr}\, \big(
       D_{\mu}U \, \overline{D}^{\mu}U^\dagger (\chi  U^\dagger - U
\chi^\dagger)\big)
 \\ &&
 -\frac{1}{3}\tilde{c}
       \,\mbox{tr}\left(D^2\chi U^\dagger- U \overline{D}^2 \chi^\dagger\right)
 \\ &&
     + \Big[ -\frac{3}{2}c^2y -cx(2+y-4xy) +\frac{1}{36}x \Big]
       \,\mbox{tr}\,\Big((\chi U^\dagger)^2-(\chi^\dagger U)^2\Big)
\\ &&  
     + \frac{2}{3}\,\tilde{c}
       \,\mbox{tr}\,\left(D_\mu U U^\dagger
        (D^\mu \chi U^\dagger+U \overline{D}^\mu \chi^\dagger)\right)
 \\ &&
     +\frac{1}{30}
       \,\mbox{tr}\, \big( F^{L\,\mu}{}_{\mu\nu}D^{\nu}U\, U^\dagger
      -F^{R\,\mu}{}_{\mu\nu} \overline{D}^{\nu}U^\dagger U \big)
 \\ &&
     - \frac{1}{36}
       \,\mbox{tr}\, \Big( \big( F^L_{\mu\nu} \big)^2
                -\big( F^R_{\mu\nu} \big)^2 \Big)
 \bigg\}
\\[0.9em] &+&
 \frac{N_c}{32\pi^2\mu^2} \bigg\{
      -\frac{4}{135}
       \Big(\partial_\mu \mbox{tr}\,\big(\chi U^\dagger - U\,\chi^\dagger
\big)\Big)
       \,\mbox{tr}\, \big( L^\mu L_\nu L^\nu \big)
 \\ &&
     +\Big[  \frac{1}{216}+\frac{3}{2}c^2y \Big]
      \Big( \partial_\mu \mbox{tr}\,\big(\chi U^\dagger - U\,\chi^\dagger \big)
      \Big)^2
 \bigg\}
\\[0.7em] &+&
 \frac{N_c}{32\pi^2\mu^2} \Big(\mbox{tr}\,
       \big(\chi U^\dagger - U\chi^\dagger \big)\Big)^2
       \bigg\{
               \Big[\frac{1}{4}c^2y-\frac{1}{3}cxy+\frac{1}{54}x \Big]
               \mbox{tr}\, (\chi U^\dagger+U\chi^\dagger)
 \\ &&
              +\frac{1}{1080} \mbox{tr}\, \big(L_\mu L^\mu \big)
       \bigg\}\,,
\end{eqnarray}
where $D^{\prime}_{\mu}*= \partial_{\mu}*+[A^L_{\mu},*]$,
$\overline{D}^{\prime}_{\mu}* = \partial_{\mu}*+[A^R_{\mu},*]$,
$F^R_{\alpha\mu\nu}=\overline{D}^{\prime}_\alpha F^R_{\mu\nu}$
and $F^L_{\alpha\mu\nu}=D^{\prime}_\alpha F^L_{\mu\nu}$.
Terms containing a factor $c=\frac{1}{6}\Big(x-\frac{1}{6y}\Big)$
are related to the $O(p^4)$ part of the field transformations
(\ref{eom4}) and those proportional to
$\tilde{c}=\frac{\pi^2}{2y}\big(\frac{1}{6}-x-y\big)^2$
are related to the transformations defined in Eqs.(\ref{soc}).

The P and C symmetries of the strong interaction allow at $O(p^6)$
structures which are proportional to
$\varepsilon_{\alpha \beta \mu \nu}$
\cite{bijnens2,leut_lec,fearing-scherer} and
which do not belong to the Wess-Zumino anomalous action. However, in
this approach these contributions disappear, because we have limited
our self to calculate only the absolute value of the quark
determinant.

\section*{\bf Conclusion}

   We have presented an effective chiral meson
Lagrangian to $O(p^6)$ in the momentum expansion,
 obtained from the bosonization of the
NJL model.
   To minimize the number of independent terms in this
expression, extensive use of the properties of covariant
derivatives and field transformations has been made.
   In contrast to previous studies of Lagrangians at $O(p^4)$, we
had to retain the next-to-leading order terms in the field
transformations
 which gave additional contributions to the
$O(p^6)$ Lagrangian in the process of transforming the
bosonized $p^4$-Lagrangian to the canonical Gasser-Leutwyler
form.

   The Lagrangian obtained at $O(p^6)$ is expected to be important in
neutral meson processes, for example,
$\eta\to\pi^0\gamma\gamma$, $\gamma\gamma \to \pi^0\pi^0$ and $K^0_L \to
\pi^0 \pi^0 \gamma$ where Born contributions from the $O(p^4)$
Lagrangian vanish \cite{ecker_lec}.
   However, taking the relevant effects correctly into account
is not a simple task since together with the results of
bosonization (i.e.\ transition to collective meson fields
through integrating out the quark degrees of freedom)
one has to address the question about the
influence of heavy vector and axial-vector resonances
\cite{reduction} and also the nonlocal corrections for the usual
local version of NJL model \cite{bilocal}.

\vspace{0.5cm}

     The authors are grateful to G.Ecker, M.L.Nekrasov and
M.K.Volkov for useful discussions and helpful comments.
   One of the authors (A.A.Bel'kov) is grateful for the hospitality
extended to him at TRIUMF, Vancouver. This work was supported in part
by a grant from the Natural Sciences and Engineering Research
Council of Canada (S.Scherer) and by the Russian Foundation for
Fundamental Research (A.A.Bel'kov and A.V.Lanyov) under grant
No.~94-02-03973.

\newpage

\begin{appendix}
\section{Identities and Field Transformations}
\subsection*{\bf Identities}

In order to reduce the number of terms as much as possible
we have made use of several identities and relations, which are listed
here.

The Lagrangian contains different types of derivatives which satisfy the
following rules:
\begin{eqnarray}
D_{\mu}(O_1O_2)&=&(D_{\mu}O_1)O_2 + O_1(\overline{D}^{\prime}_{\mu}O_2)
                =(D_{\mu}^{\prime}O_1)O_2 + O_1(D_{\mu}O_2)\,,
\nonumber \\
\overline{D}_{\mu}(O_1O_2)&=&(\overline{D}_{\mu} O_1) O_2
                            +O_1(D^{\prime}_{\mu} O_2)
                           = (\overline{D}^{\prime}_{\mu} O_1) O_2
                            +O_1(\overline{D}_\mu O_2  )\,,
\nonumber \\
D^{\prime}_{\mu}(O_1O_2)&=&(D^{\prime}_{\mu}O_1)O_2 +
                           O_1(D^{\prime}_{\mu}O_2)
                =(D_{\mu}O_1)O_2 + O_1(\overline{D}_{\mu}O_2)\,,
\nonumber \\
\overline{D}^{\prime}_{\mu}(O_1O_2)&=&
                             (\overline{D}^{\prime}_{\mu} O_1) O_2
                            +O_1(\overline{D}^{\prime}_{\mu} O_2)
                           = (\overline{D}_{\mu} O_1) O_2
                            +O_1 (D_\mu O_2  )\,.
\label{appendix1}
\end{eqnarray}
In order to reduce the number of terms which contain Goldstone bosons only,
we applied the following relations
\begin{eqnarray}
{} [D_{\mu},D_{\nu}] O = F^L_{\mu \nu}O - OF^R_{\mu \nu}\;\;,\;\;
&& [\overline{D}_{\mu}, \overline{D}_{\nu}]O =
                         F^R_{\mu \nu} O - O F^L_{\mu \nu}\,,
\nonumber \\
{}  [D^{\prime}_{\mu}, D^{\prime}_{\nu}]  O =  [F^L_{\mu \nu}, O]\;\;,\;\;
&&    [\overline{D}^{\prime}_{\mu}, \overline{D}^{\prime}_{\nu}]O =
                                             [F^R_{\mu \nu}, O]\,.
\label{appendix2}
\end{eqnarray}
We have also made use relations arising from the unitarity of the
matrix U:
$$
   D_{\mu}U\,U^\dagger = -U\,\overline{D}_{\mu}U^\dagger\;\;,\;\;
 D_{\mu}D_{\nu}U\,U^\dagger +
      U\,\overline{D}_{\mu}\overline{D}_{\nu}U^\dagger =
-\left(
        D_{\mu}U\,\overline{D}_{\nu}U^\dagger
       +D_{\nu}U\,\overline{D}_{\mu}U^\dagger
 \right) \,.
$$

\subsection*{\bf Field Transformations}

The initial Lagrangian also contained redundant terms which can
be eliminate with the help of field transformations.
At $O(p^4)$ the use of field transformations and a naive application
of the classical EOM are equivalent.
At the next order in the momentum expansion, $O(p^6)$, the method of
field transformations will give rise to
contributions which one would miss using the classical EOM only.
Before the application of field transformations
the most general Lagrangian of $O(p^4)$ typically has the form (see
Eq.(\ref{ap2}))
\footnote{Note our different convention for
the definitions of the tensors $F^{L,R}$: $F^{L,R}=-iF^{R,L}_{G\&L}$.}
\begin{eqnarray}
\lefteqn{{\cal L}_4  =
  L_1' \left ( \mbox{tr}\,(D_{\mu}U \overline{D}^{\mu}U^{\dagger}) \right)^2
+ L_2' \mbox{tr}\, \left (D_{\mu}U \overline{D}_{\nu}U^{\dagger}\right)
  \mbox{tr}\, \left (D^{\mu}U \overline{D}^{\nu}U^{\dagger}\right)}
\nonumber \\  &&
+ L_3' \mbox{tr}\, \left (D_{\mu}U \overline{D}^{\mu}U^{\dagger}D_{\nu}U
                     \overline{D}^{\nu}U^{\dagger}  \right )
+ L_4' \mbox{tr}\, \left ( D_{\mu}U \overline{D}^{\mu}U^{\dagger} \right )
       \mbox{tr}\, \left( \chi U^{\dagger}+ U \chi^{\dagger} \right )
\nonumber \\ &&
+ L_5' \mbox{tr}\, \left( D_{\mu}U \overline{D}^{\mu}U^{\dagger}
                 (\chi U^{\dagger}+ U \chi^{\dagger})\right)
+ L_6' \left( \mbox{tr}\, \left ( \chi U^{\dagger}+ U \chi^{\dagger} \right)
\right)^2
\nonumber \\ &&
+ L_7' \left( \mbox{tr}\, \left ( \chi U^{\dagger} - U \chi^{\dagger} \right)
\right)^2
+ L_8' \mbox{tr}\, \left ( U \chi^{\dagger} U \chi^{\dagger}
                + \chi U^{\dagger} \chi U^{\dagger} \right )
\nonumber \\ &&
+ L_9' \mbox{tr}\, \left ( F^L_{\mu\nu} D^{\mu} U \overline{D}^{\nu}
U^{\dagger}
                + F^R_{\mu\nu} \overline{D}^{\mu} U^{\dagger} D^{\nu} U \right)
- L_{10}' \mbox{tr}\, \left ( U F^R_{\mu\nu} U^{\dagger} F_L^{\mu\nu} \right)
\nonumber \\ &&
- H_1' \mbox{tr}\, \left( F^R_{\mu\nu} F^{\mu\nu}_R + F^L_{\mu\nu}F^{\mu\nu}_L
\right)
+ H_2' \mbox{tr}\, \left ( \chi \chi^{\dagger} \right )
\nonumber\\ &&
+ \lambda_1 \mbox{tr}\,\left( D^2U \overline{D}^2 U^\dagger \right)
+ \lambda_2 \mbox{tr}\,\left( D^2 U\chi^\dagger + \chi \overline{D}^2
U^\dagger\right).
\label{l41}
\end{eqnarray}
It contains 2 more structures than the standard Lagrangian of Gasser
and Leutwyler.
In order to eliminate the two additional terms, one rewrites
$D^2 U\,U^\dagger$ and $U \overline{D}^2 U^\dagger$
applying the chain rule to $U U^\dagger=1$.
After some algebra the Lagrangian of Eq.(\ref{l41}) can be written as
 \begin{equation}
 {{\cal L}}_4  =  {{\cal L}}_4^{G \& L}
 +c_1 \mbox{tr}\,\left((D^2U U^\dagger-U D^2U^\dagger){\cal
O}^{(2)}_{EOM}\right)
 +c_2 \mbox{tr}\,\left((\chi U^\dagger-U\chi^\dagger){\cal
O}^{(2)}_{EOM}\right),
 \label{l42}
\end{equation}
where ${\cal L}_4^{G\& L}$ is the Gasser and Leutwyler Lagrangian
defined in \cite{gasser} and
${\cal O}^{(2)}_{EOM}$ has the functional form of the classical EOM of
$O(p^2)$:
\begin{equation}
 {\cal O}^{(2)}_{EOM}(U)=D^2 U U^\dagger-U \overline{D}^2 U^\dagger
 -\chi U^\dagger +U\chi^\dagger
 +\frac{1}{3}\mbox{tr}\,\left(\chi U^{\dagger}-U\chi^{\dagger}\right).
 \label{eom2}
 \end{equation}
The unprimed (G\&L) and primed coefficients are related through
 \begin{eqnarray}
 \label{coeffrel}
 &&
 L_1=L_1',\quad L_2=L_2',\quad L_3=L_3'+\lambda_1,\quad L_4=L_4',\quad
 L_5=L_5'-\lambda_2,\nonumber\\
 &&
 L_6=L_6',\quad L_7=L_7'+\frac{\lambda_1}{12}+\frac{\lambda_2}{6},\quad
 L_8=L_8'-\frac{\lambda_1}{4}-\frac{\lambda_2}{2},\quad
 L_9=L_9',\nonumber\\
 &&
 L_{10}=L_{10}',\quad H_1=H_1',\quad
 H_2=H_2'+\frac{\lambda_1}{2}+\lambda_2,\nonumber\\
 &&
 c_1=-\frac{\lambda_1}{4},\quad
 c_2=-\frac{\lambda_1}{4}-\frac{\lambda_2}{2}.
 \end{eqnarray}
 Using the field transformation technique \cite{fields} we will get rid of
 the last two terms of Eq.\ (\ref{l42}).
 For that purpose we write
 \begin{equation}
 \label{uvrel}
 U(x)=\exp(iS_2(V))V(x),
 \end{equation}
 where $S_2(V)$ is given by
\begin{eqnarray}
 S_2(V)&=&-i\frac{\lambda_1}{F_0^2} (D^2VV^{\dagger}-V\overline{D}^2
 V^{\dagger})
\nonumber \\ &&
 -i \bigg( \frac{\lambda_1}{F_0^2} + \frac{2\lambda_2}{F_0^2}\bigg)
 \left(\chi V^{\dagger}-V \chi^{\dagger}-\frac{1}{3}
 Tr(\chi V^{\dagger}-V \chi^{\dagger})\right)\,.
\label{s2}
\end{eqnarray}
If we insert $U=\exp(iS)V$ into ${\cal L}_2(U)$ we obtain
 \begin{equation}
 {\cal L}_2(U)={\cal L}_2(V)+\delta^{(1)}{\cal L}_2(V,S)+
 \delta^{(2)}{\cal L}_2(V,S)+\dots
 \label{l2v}
 \end{equation}
 In Eq.\ (\ref{l2v}) we have dropped an irrelevant total derivative.
 The superscripts denote the power of S (or $D_\mu S, \dots$)
 and the corresponding expressions are given by
\begin{eqnarray}
 \delta^{(1)}{\cal L}_2(V,S_2)&=&\frac{F^2_0}{4}
 \mbox{tr}\,\left(iS_2{\cal O}^{(2)}_{EOM}(V)\right)=O(p^4),
\nonumber\\
 \delta^{(2)}{\cal L}_2(V,S_2)&=&\frac{F^2_0}{4}\mbox{tr}\,\left( S_2(D_\mu V
V^\dagger
 D^\mu S_2 - D^\mu S_2 D_\mu V V^\dagger - D^2 S_2) \right.
\nonumber\\  &&
\left.-\frac{1}{2}(\chi V^\dagger+V \chi^\dagger)
 S_2^2\right)=O(p^6),
\nonumber\\
 \delta^{(3)}{\cal L}_2(V,S_2)&=&O(p^2)\times O(S_2^3)=O(p^8).
\label{dell}
\end{eqnarray}
 The last term is only interesting at $O(p^8)$ and thus we do not give its
 explicit form.

 With our choice of $S_2$, eq.(\ref{s2}),
 the term $\delta^{(1)}{\cal L}_2(V,S_2)$ precisely cancels the last two
 contributions of Eq.\ (\ref{l42}) ($U\rightarrow V$ at  $O(p^4)$;
 we postpone the discussion of the change in ${{\cal L}}_4$).
 Let us now investigate $\delta^{(2)}{\cal L}_2(V,S_2)$. For that purpose
 we define the following operators
 \begin{eqnarray}
 \label{aops}
 A_1&=&Tr\left((\chi U^\dagger-U \chi^\dagger)\{L_\mu,
 (D^\mu \chi U^\dagger+U \overline{D}^\mu
\chi^\dagger)\}\right),\nonumber\\
 A_2&=&Tr\left(\chi U^\dagger-U \chi^\dagger\right)
 Tr\left(L_\mu(D^\mu \chi U^\dagger+U \overline{D}^\mu
\chi^\dagger)\right),
 \nonumber\\
 A_3&=&Tr\left((\chi U^\dagger-U \chi^\dagger)L_\mu
 (\chi U^\dagger-U \chi^\dagger)D^\mu U U^\dagger\right),\nonumber\\
 A_4&=&Tr\left(\chi U^\dagger-U \chi^\dagger\right)
 Tr\left(D_\mu U D^\mu U^\dagger (\chi U^\dagger - U
\chi^\dagger)\right),
 \nonumber\\
 A_5&=&Tr\left((\chi U^\dagger-U\chi^\dagger)(\overline{D}^2\chi
U^\dagger -
 U D^2\chi^\dagger)\right),\nonumber\\
 A_6&=&Tr\left(\chi U^\dagger - U \chi^\dagger\right)
 Tr\left(D^2\chi U^\dagger- U \overline{D}^2 \chi^\dagger\right).
 \end{eqnarray}
 In terms of the structures of Eq.\ (\ref{aops}) the second order (in $S_2$)
 change reads
 \begin{equation}
 \frac{(\lambda_1+\lambda_2)^2}{F_0^2}
\Big(-A_1+\frac{2}{3}A_2+A_3+\frac{1}{3}A_4
 +A_5-\frac{1}{3}A_6\Big).
 \label{soc}
 \end{equation}
In our NJL-based approach the coefficients are
$\lambda_1=\frac{1}{6}\frac{N_c}{16\pi^2}$ and
$\lambda_2=\frac{N_c}{16\pi^2}(-x-y)$.

 When deriving Eq.\ (\ref{soc}) we extensively made use of the total derivative
 argument and also field transformations where the generators are of
 $O(p^4)$. (We have been somewhat sloppy in the sense that we used the symbol
$U$
 again for the final expression. Of course it is {\em not} the same
interpolating
 field we start off with.)

The modification of ${\cal L}_4$ has a similar form as that of ${\cal L}_2$
in Eq.(\ref{l2v})
\begin{equation}
{\cal L}_4(U)={\cal L}_4(V)+\delta^{(1)}{\cal L}_4(V,S)+ O(p^8)
\label{l4v}
\end{equation}
where
\begin{equation}
\delta^{(1)}{\cal L}_4(V,S)=\frac{F^2_0}{4}
 \mbox{tr}\,\left(iS{\cal O}^{(4)}_{EOM}(V)\right)=O(p^6)
\label{eom4}
\end{equation}
{}From the Lagrangian in Eq.(\ref{l41}) we obtain for the $O(p^4)$
contribution to the EOM operator
\begin{equation}
{\cal O}^{(4)}_{EOM}(U)=\frac{4}{F_0^2}\Big(E_4 - \frac{1}{3}tr(E_4)\Big)\,,
\label{eomgen}
\end{equation}
where
\begin{eqnarray}
E_4 &=& \bigg(2L_1^\prime - L_2^\prime \bigg)
        \mbox{tr}\,\big( D_\mu U \, \overline{D}^\mu U^\dagger \big)
        \cdot \big(D^2 U \, U^\dagger -U\,\overline{D}^2 U^\dagger\big)
\nonumber \\
    &+& 2L_2^\prime
        \Big[
        -U\,\overline{D}_\mu \left(\overline{D}_\nu U^\dagger\,D^\mu
         U\,\overline{D}^\nu U^\dagger\right)
        +D_\mu \left(D_\nu U\,\overline{D}^\mu U^\dagger\,D^\nu
         U\right) U^\dagger
        \Big]
\nonumber \\
    &+&(4L_2^\prime +2L_3^\prime )
            \Big[
                -U\overline{D}_\mu
                 \left( \overline{D}^\mu U^\dagger\,D_\nu U\,\overline{D}^\nu
                      U^\dagger \right)
                +D_\mu \left( D_\nu U\,\overline{D}^\nu
                              U^\dagger\,D^\mu U \right)\, U^\dagger
            \Big]
\nonumber \\
    &+& L_4^\prime \Big[
       \mbox{tr}\,\big(\chi U^\dagger + U \chi^\dagger \big)
       \cdot \big(D^2 U \, U^\dagger - U\,\overline{D}^2 U^\dagger \big)
     + \mbox{tr}\,\big( D_\mu U \, \overline{D}^\mu U^\dagger \big)
       \cdot \big(U \chi^\dagger -\chi U^\dagger \big) \Big]
\nonumber \\
    &+& L_5^\prime
        \Big[
         -  U\,\overline{D}_\mu\left(\overline{D}^\mu
            U^\dagger\left(\chi U^\dagger+U \chi^\dagger\right)\right)
         +  D_\mu \left( \left(\chi U^\dagger+U\chi^\dagger\right) D^\mu
U\right)
            U^\dagger
\nonumber \\ &&  \quad\quad \;\;
         +  \,U\chi^\dagger\,D_\mu U \, \overline{D}^\mu U^\dagger
         -  \,D_\mu U \, \overline{D}^\mu U^\dagger \, \chi U^\dagger
        \Big]
\nonumber \\
    &+& 2 L_6^\prime \mbox{tr}\,\big(\chi U^\dagger+U \chi^\dagger \big)
        \cdot \big(U \chi^\dagger -\chi U^\dagger \big)
\nonumber \\
    &-& 2 L_7^\prime \mbox{tr}\,\big(\chi U^\dagger -U \chi^\dagger \big)
        \cdot \big(U \chi^\dagger +\chi U^\dagger \big)
\nonumber \\
    &+& L_8^\prime
            \Big[
                 \left(U \chi^\dagger \right)^2 -\left( \chi U^\dagger
\right)^2
            \Big]
\nonumber \\
    &+& L_9^\prime
            \Big[
                 -U\,\overline{D}^\nu (F^R_{\mu\nu}\,\overline{D}^\mu U^\dagger
)
                 +D^\mu (D^\nu U\, F^R_{\mu\nu}) U^\dagger
                 -U\overline{D}^\mu ( \overline{D}^\nu U^\dagger\,F^L_{\mu\nu})
                 +D^\nu (F^L_{\mu\nu} D^\mu U )U^\dagger
            \Big]
\nonumber \\
    &-& L_{10}^\prime
            \Big[
                 U\,F^R_{\mu\nu}\,U^\dagger\,F^{L\,\mu\nu}
                -F^L_{\mu\nu}\,U\,F^{R\,\mu\nu}\,U^\dagger
            \Big]
\nonumber \\
    &+& \lambda_1
            \Big[ U\,\overline{D}^2\overline{D}^2 U^\dagger
                 -D^2 D^2 U \, U^\dagger \Big]
\nonumber \\
    &+& \lambda_2
           \Big[ U\,\overline{D}^2 \chi^\dagger -D^2 \chi \, U^\dagger \Big]\,.
\label{eom42}
\end{eqnarray}
\end{appendix}

%
%

\end{document}